\documentstyle[11pt,paspconf]{article} 
\input epsf

\begin{document}

\title{H$\alpha$ from HVCs}
\author{S. L. Tufte, R. J. Reynolds, and L. M. Haffner}
\affil{Department of Astronomy, University of Wisconsin -- Madison, 475 N. 
        Charter St., Madison, WI 53706}

\begin{abstract}
Optical emission lines provide an important new window on the HVCs.
Recent studies of the H$\alpha$ line reveal that ionized hydrogen is
pervasively associated with the neutral hydrogen in HVCs.  The
Wisconsin H$\alpha$ Mapper (WHAM) instrument has so far detected
H$\alpha$ from high-velocity clouds in the M, A, and C complexes.  We
find a close spatial correspondence between the neutral and ionized
portions of the HVCs with some evidence that the ionized gas envelopes
the neutral part of the clouds.  The velocities of the H$\alpha$ and
21-cm detections are well correlated, but the intensities are not.  If
the clouds are photoionized, the H$\alpha$ intensity is a direct
measure of the Lyman continuum flux in the Galactic halo.  Forthcoming
observations of the H$\alpha$ line in combination with other emission
lines will give new insights into the high-velocity cloud phenomena
and will also probe the physical conditions of their environment.
\end{abstract}

\section{Introduction}

The study of H$\alpha$ and other optical emission lines from HVCs,
just now becoming technologically viable, holds great promise for
broadening our understanding of both HVCs and their environments.  The
H$\alpha$ line gives information about the ionized hydrogen associated
with the HVC, specifically, the emission measure, and the spatial and
kinematic relationship to the neutral gas, already well studied
through the 21-cm line.  If the H$\alpha$ emitting regions are
photoionized, then the H$\alpha$ intensity is a direct measure of the
Lyman continuum flux incident upon the cloud.  A knowledge of how this
flux varies with distance above the Galactic plane and with
Galactocentric radius can have an important bearing on more general
questions regarding ionized gas in the disk and halo of the Milky Way.
If, on the other hand, the ionizations producing the H$\alpha$ photons
are due to collisional processes induced by the rapid motion of the
HVCs (i.e., shocks), then the intensity is a measure of the ambient
density through which the cloud is moving.  The extent of the low
density Galactic halo has important implications for considerations of
dark matter in the Galaxy as well as for properly interpreting QSO
absorption line measurements (e.g., Sembach et al.  1995).
Measurements of other emission lines such as [O~III] $\lambda$5007,
[S~II] $\lambda$6716, and [N~II] $\lambda$6584 should distinguish
between these two ionization mechanisms.  In the near future, it will
be possible to map entire HVC complexes in H$\alpha$, allowing
detailed comparisons of the distribution of neutral and ionized gas in
the clouds, as well as providing the opportunity to look for clues
that connect HVCs to features in the ISM at lower velocities.

The extreme faintness of the H$\alpha$ emission from HVCs made it
nearly impossible to study in the past.  Reynolds (1987) searched for
H$\alpha$ in six directions and could only place upper limits ranging
from 0.2 R to 0.6 R.  One Rayleigh (R) is 10$^{6}$ / 4$\pi$ photons
cm$^{-2}$ s$^{-1}$ sr$^{-1}$ or 2.41 $\times$ 10$^{-7}$ ergs cm$^{-2}$
s$^{-1}$ sr$^{-1}$ at H$\alpha$.  Kutyrev and Reynolds (1989)
successfully detected H$\alpha$ (I$_\alpha$ = 0.08 $\pm$ 0.02 R) from
a very high velocity cloud (v$_{lsr}$ = -300 km s$^{-1}$) in Cetus.
M\"{u}nch \& Pitz (1989) detected an intensity I$_\alpha$ = 0.15 R
toward the M~II cloud, a high velocity cloud in the M complex.
Songaila et al. (1989) claimed a detection of H$\alpha$ from complex C
at the 0.03 R level; however, Tufte et al. (1998) have observed the
same direction and report a significantly higher intensity (see
below).  In general, these observations involved long integration
times and produced null or marginal results.

The advent of CCD detectors with both high quantum efficiency and
extremely low dark noise has dramatically improved the prospects for
optical emission line studies of HVCs.  Coupling these detectors to
Fabry-Perot based spectrometers has made such optical observations a
viable and promising field of study.  Consider for example, the recent
Weiner \& Williams (1996) study of H$\alpha$ emission towards the
Magellanic Stream using the Rutgers Imaging Fabry-Perot system coupled
to the 1.5~m CTIO telescope.  They detected H$\alpha$ emission towards
5 out of their 7 observation directions and found that the H$\alpha$
``bright'' spots ($\simeq$ 0.2 -- 0.4~R) appear to be correlated with
H~I cloud edges.  Weiner \& Williams argue that the H$\alpha$ arises
from gas ionized by ram pressure heating by low density (n$_{H}$
$\sim$ 10$^{-4}$) gas in the Galactic halo, but this conclusion is a
matter of controversy (see Bland-Hawthorn \& Maloney 1998).

\section{WHAM Observations of H$\alpha$ from HVCs}
The Wisconsin H-Alpha Mapper (WHAM) is a unique new instrument for the
detection and study of faint optical emission lines from diffuse
ionized gas in the disk and halo of the Galaxy (Reynolds et al.  1998;
Tufte 1997).  Its characteristics make it very well suited to the
study of emission lines from HVCs and have led to the detection of
optical line emission from a number of HVCs (Tufte 1997; Tufte et
al. 1998).

WHAM consists of a 15 cm aperture, dual-etalon Fabry-Perot
spectrometer, which provides much higher throughput at a given
spectral resolution than grating spectrometers.  The high spectral
resolution (12 km~s$^{-1}$ FWHM) is well matched to the 20--25
km~s$^{-1}$ line widths (21-cm) typical of HVCs and makes possible
studies of the kinematics of the emitting gas.  The spectrometer
is coupled to a dedicated 0.6 m telescope that produces a 1$^{\rm o}$
diameter beam on the sky, well matched to the angular sizes of HVCs.
The large etalons combined with a high efficiency CCD detector
provides unsurpassed sensitivity, which results in clear (10$\sigma$) 
detections of emissions with an intensity of 0.1 R, typical of 
the H$\alpha$ from HVCs (see below), in 30 minutes integration time.
Finally, WHAM is tunable to any wavelength between 4800 $\AA$ and 
7200 $\AA$, and therefore can be used to study a variety of 
lines from HVCs.

\subsection{H$\alpha$ from the M Complex}
\begin{figure}
  \centerline{
  \epsfysize = 5.0in
  \epsffile{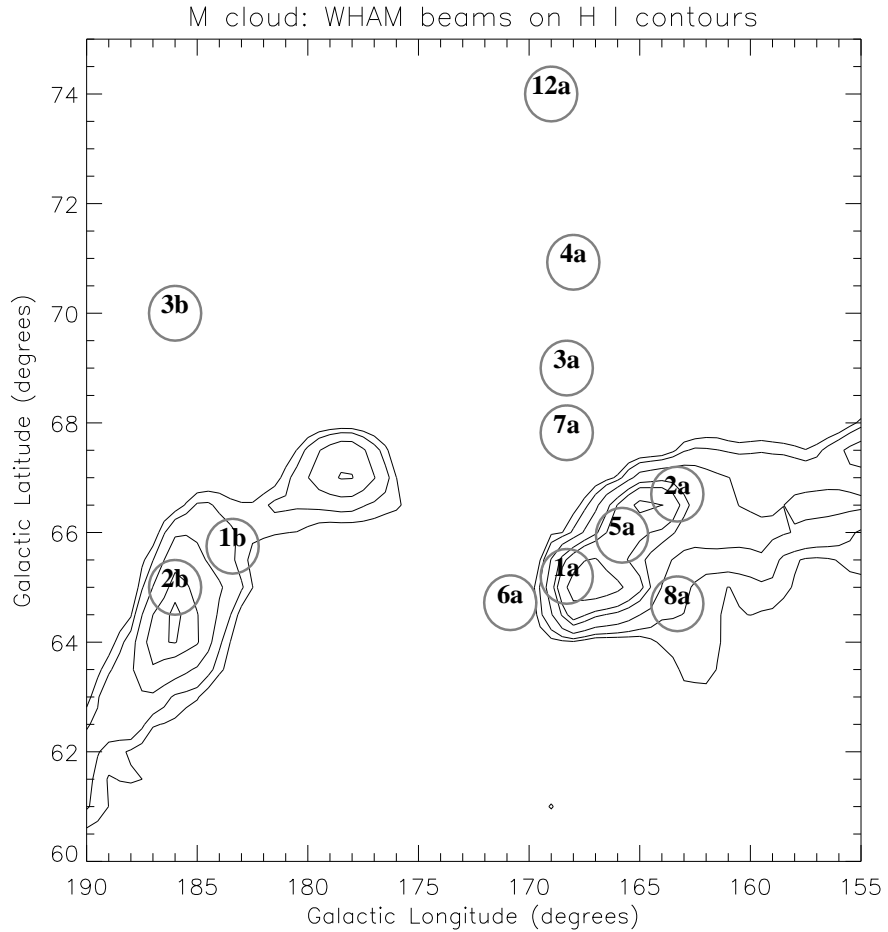} }
  \caption{WHAM look directions for the M cloud superposed on 21-cm contours. 
  The contour levels correspond to N$_{\rm H I}$ = 2.0, 3.0, 5.0, 8.0, 10.0,
  and 15.0 $\times$ 10$^{19}$ cm$^{-2}$.}
\label{fig4-map_m}
\end{figure}

The WHAM H$\alpha$ observation directions for the M complex region
are shown in Figure~\ref{fig4-map_m} superposed on 21-cm contours from
the Leiden/Dwingeloo survey (Hartmann \& Burton
1997).  The 21-cm contours indicate the
distribution of neutral hydrogen at high velocity (v$_{lsr}$ $\leq$
-80 km s$^{-1}$).

\begin{figure}[t]
  \centerline{\hbox{
      \epsfysize=2.7in \epsfbox{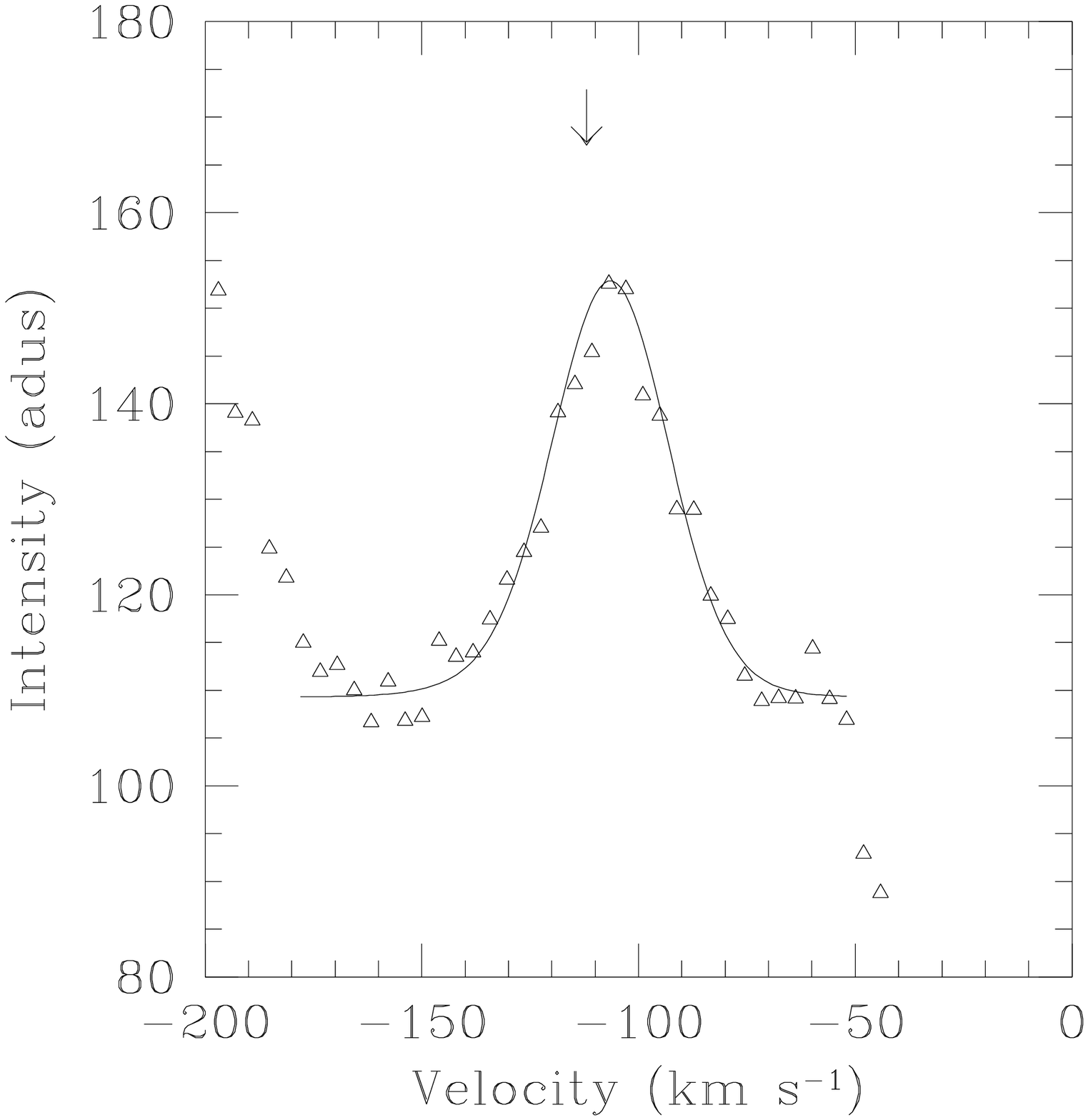}
      \epsfysize=2.7in \epsfbox{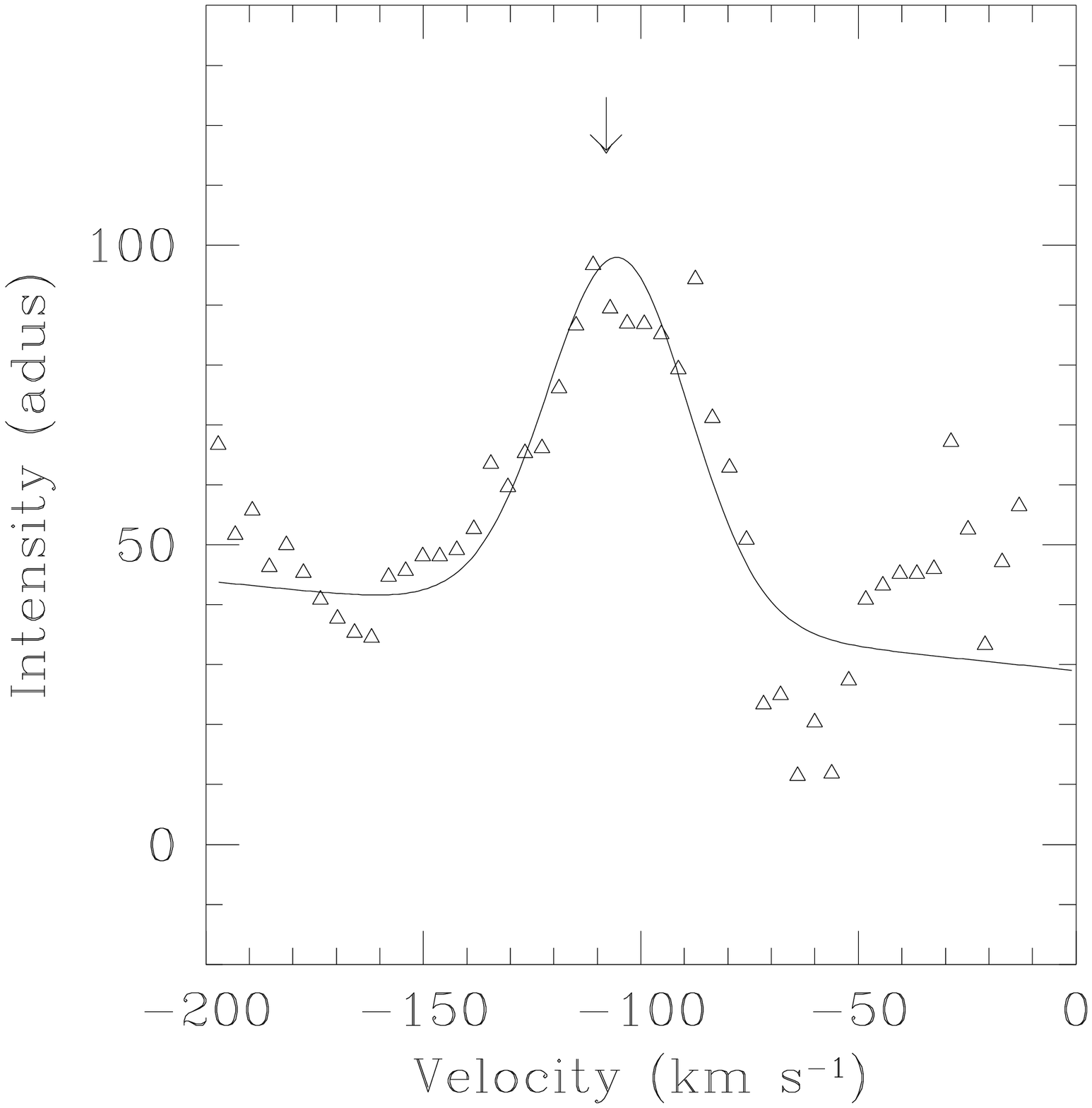}}
    }
  \caption{H$\alpha$ from the M I cloud: The left
  spectrum is from the direction $l$ = 168.3$^{\rm o}$, $b$ =
  65.2$^{\rm o}$ (1a in Fig.~\ref{fig4-map_m}), which is centered on
  the 21-cm enhancement identified as the M~I cloud.  The right
  spectrum is toward the direction $l$ = 170.85$^{\rm o}$, $b$ =
  64.72$^{\rm o}$ (6a), which is just off the M~I cloud where the
  21-cm intensity is much lower.  The arrow in each case shows the
  velocity of the 21-cm HVC detection.  These spectra are actually
  produced by subtracting the spectra in ``off'' cloud directions from
  the spectra obtained toward the clouds.  The left spectrum is 1a -
  4a, the right is 6a - 12a (see Fig.~\ref{fig4-map_m}).}
\label{fig-hvcs}
\end{figure}

Figure~\ref{fig-hvcs} shows two of the WHAM H$\alpha$ spectra toward
the M complex, formed by subtracting the spectrum of a nearby ``off''
direction, free from high velocity gas in the 21-cm maps, from a
spectrum taken ``on'' the direction of interest.  This very powerful
technique eliminates night sky features and other systematic
uncertainties that would otherwise obscure the very weak emission from
the HVC and isolates the H$\alpha$ emission associated with the HVC.
The left spectrum is toward $l$ = 168.3$^{\rm o}$, $b$ = 65.2$^{\rm
o}$ (direction 1a in Fig.~\ref{fig4-map_m}), which is centered on the
peak of the 21-cm contours defining the M~I cloud.  An emission line
is clearly detected at a velocity with respect to the local standard
of rest v$_{lsr}$ = $-106$ km~s$^{-1}$.  A best fit gaussian to this
spectrum (solid curve) implies an intensity of 0.078 R and a width
(FWHM) of 27 km~s$^{-1}$ for this emission line.  The arrow in the
figure denotes the velocity of the 21-cm emission line associated with
the neutral hydrogen.  The velocities of the emission from the neutral
and ionized gas agree to within the measurement uncertainties.

The right spectrum in Figure~\ref{fig-hvcs} was taken toward the
nearby direction $l$ = 170.85$^{\rm o}$, $b$ = 64.72$^{\rm o}$
(direction 6a in Fig.~\ref{fig4-map_m}).  Once again an emission line
is clearly detected.  The fit implies an H$\alpha$ intensity of 0.20 R
at a velocity of v$_{lsr}$ = $-105$ km~s$^{-1}$.  This direction is
just off the M~I cloud 21-cm enhancement in a location where the H~I
column density is less than 1/10th that in the first direction (1a).
Nevertheless, the H$\alpha$ emission is approximately twice that in
the first direction.  This implies that there is not a one-to-one
correspondence between 21-cm and H$\alpha$ emission in this HVC, and
that in some parts of the cloud ionized gas dominates.  The
implications of this result will be considered further in
Section~\ref{sec-disc}.

\subsection{H$\alpha$ from the A and C Complexes}

\begin{figure}
  \centerline{\hbox{
      \epsfysize=2.7in \epsfbox{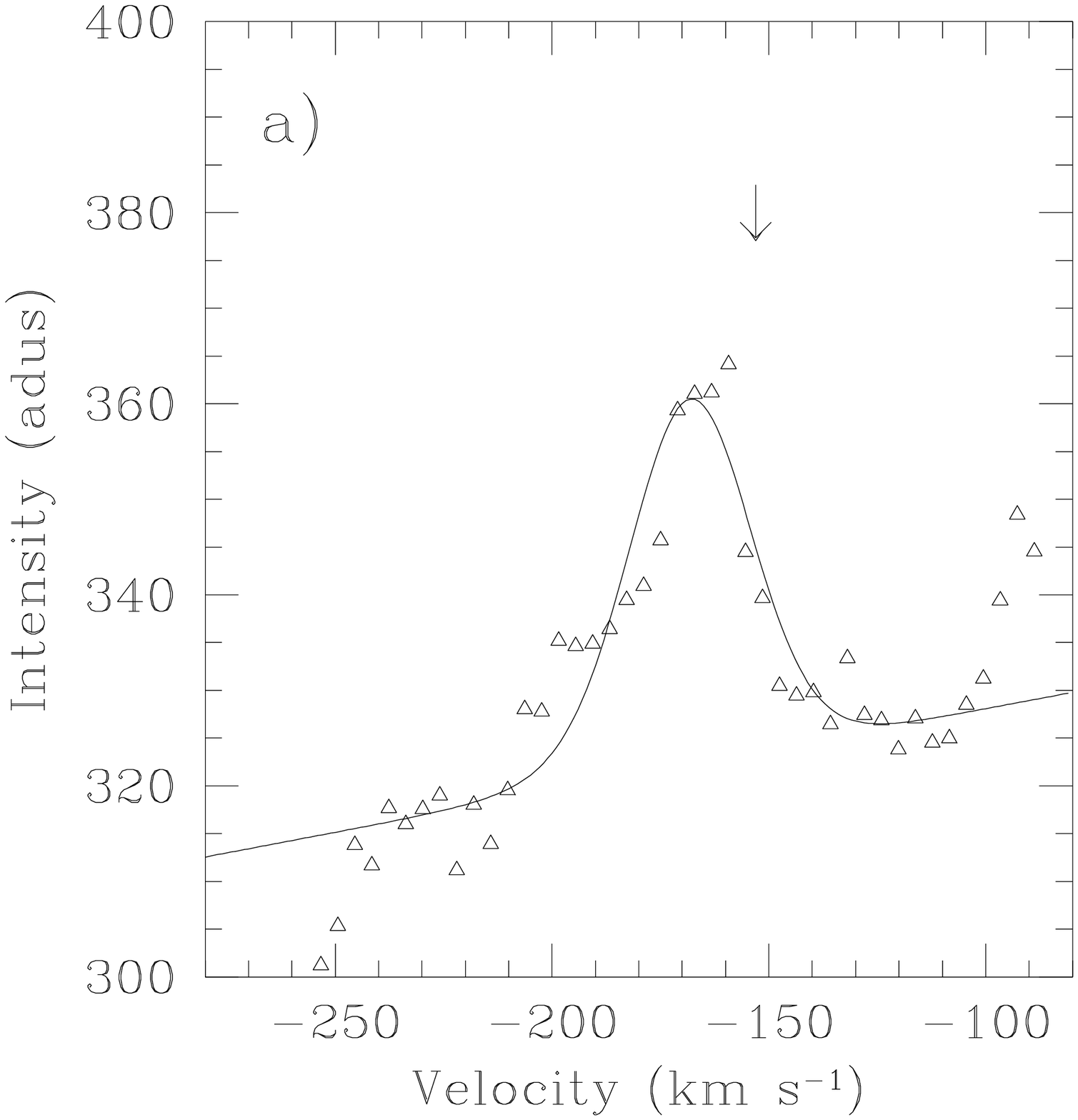}
      \epsfysize=2.7in \epsfbox{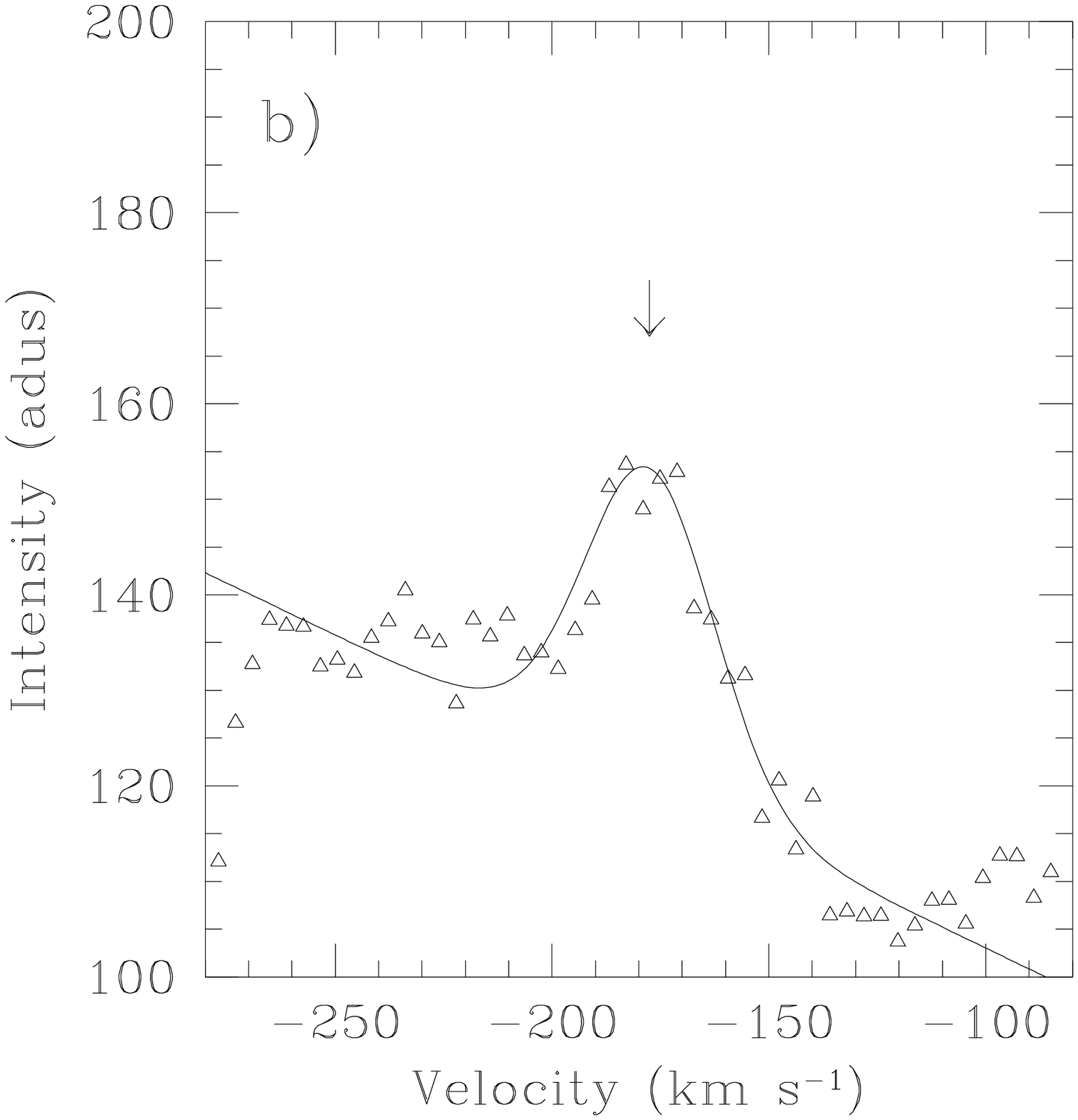}}
    }
  \caption{H$\alpha$ from the A Complex.  a) A III cloud at $l$ =
  148.5$^{\rm o}$, $b$ = 34.5$^{\rm o}$.  b) A IV cloud at $l$ =
  153.6$^{\rm o}$, $b$ = 38.2$^{\rm o}$.  Both are normalized to 900s
  exposure.}
\label{fig-hvc_a3}
\end{figure}

H$\alpha$ spectra for the A Complex are shown in
Figure~\ref{fig-hvc_a3}.  Notice that the spectrometer was ``tuned''
to cover a higher negative velocity interval than for the M complex
observations.  This is accomplished by decreasing the optical path
length in the Fabry-Perot etalon gaps by changing the gas pressure in
the etalon chambers.  The spectra show clear detections of H$\alpha$
emission from both the A~III and A~IV clouds at velocities (LSR) of
-167 $\pm$ 1 km~s$^{-1}$ and -178 $\pm$ 1 km~s$^{-1}$, respectively.
The arrows show the velocity of the 21-cm line in the
Leiden/Dwingeloo survey data.  The H$\alpha$ intensities derived from
the fits are I$_{\alpha}$ = 0.08 $\pm$ 0.01 R\@, and 0.09 $\pm$ 0.01
R, for the A~III and A~IV clouds, respectively.

\begin{figure}[t]
  \centerline{
  \epsfysize = 4.0in
  \epsffile{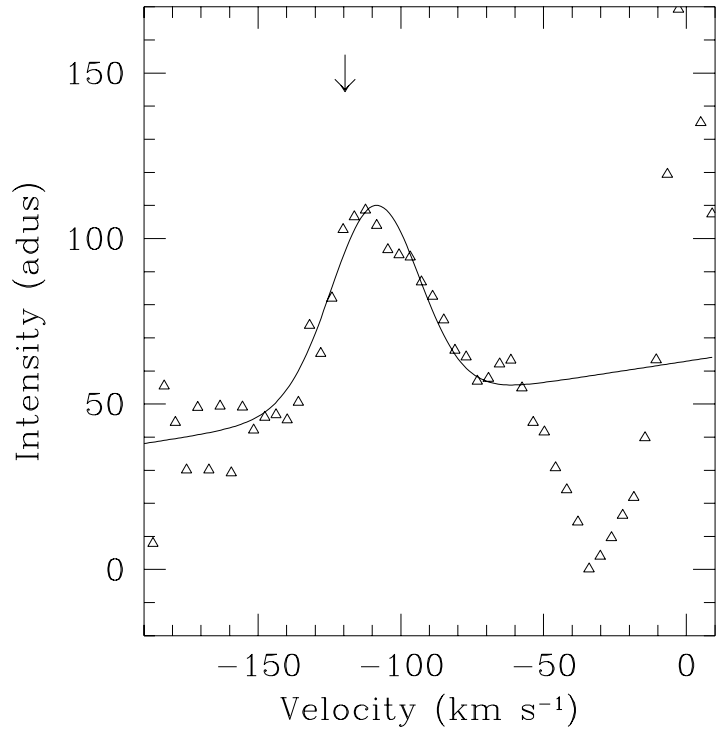} }
  \caption{H$\alpha$ from the C cloud at $l$ = 84.3$^{\rm o}$, 
  $b$ = +43.7$^{\rm o}$ (normalized to 900s exposure).}
\label{fig-hvc_c}
\end{figure}

Figure~\ref{fig-hvc_c} shows the H$\alpha$ difference spectrum for the
C complex direction ($l$ = 84.3$^{\rm o}$, $b$ = +43.7$^{\rm o}$).  While
the baseline is not well determined, particularly on the red side of
the spectrum (see Section~\ref{sec-lockman} below), an H$\alpha$
emission component is clearly detected at a velocity matching that of
the 21-cm component.  The intensity of this component is I$_{\alpha}$
= 0.13 $\pm$ 0.03 R, and the velocity is v$_{lsr}$ = -111 $\pm$ 2
km~s$^{-1}$.  This higher signal-to-noise observation is inconsistent
with the results of Songaila, Bryant, \& Cowie
(1989), who observed this same direction (see
Position 1 in their Fig. 1a) and reported detecting H$\alpha$ at an
intensity of 0.03 R with an estimated uncertainty of 50\%.
Table~\ref{tab-obs_results} summarizes the WHAM results for the M, A,
and C complexes.

\begin{table}[htb]
\begin{center}
\caption{H$\alpha$ and 21-cm Emission from HVCs}
\begin{tabular}{rccccc} & & & & & \\ \hline \hline
 & & & & & \\
 & \multicolumn{2}{c}{H$\alpha$} &  
& \multicolumn{2}{c}{21-cm} \\ 
\cline{2-3} \cline{5-6} \\
Name & I$_{H\alpha}$ & V$_{H\alpha}$ & & I$_{21cm}$ & 
V$_{21cm}$ \\
 & (R) & (km s$^{-1}$) & & (10$^{20}$ cm$^{-6}$ pc) & 
(km s$^{-1}$) \\ \hline
 & & & & & \\

M I cloud & & & & & \\ 
1a-4a  & 0.08 $\pm$ 0.01 & -106 $\pm$ 1 & & 1.22 & -113 \\
2a-4a  & 0.20 $\pm$ 0.02 & -95  $\pm$ 1 & & 1.17 & -101 \\
5a-4a  & 0.06 $\pm$ 0.02 & -103 $\pm$ 3 & & 1.04 & -116 \\
6a-12a & 0.18 $\pm$ 0.02 & -105 $\pm$ 1 & & --- & --- \\
8a-4a  & 0.11 $\pm$ 0.02 & -92  $\pm$ 2 & & 0.35 & -112 \\
 & & & & & \\ 
M II cloud & & & & & \\ 
1b-4a  & 0.12 $\pm$ 0.01 & -61  $\pm$ 2 & & 1.33 & -72  \\
2b-4a  & 0.16 $\pm$ 0.03 & -78  $\pm$ 2 & & 1.50 & -80  \\ 
 & & & & & \\ 
A cloud & & & & & \\ 
A III  & 0.08 $\pm$ 0.01 & -167 $\pm$ 1 & & 1.40 & -153 \\
A IV   & 0.09 $\pm$ 0.01 & -178 $\pm$ 1 & & 1.34 & -177 \\
 & & & & & \\ 
C cloud & & & & & \\ 
C      & 0.13 $\pm$ 0.03 & -111 $\pm$ 2 & & 0.54 & -120 \\
\hline
\end{tabular}
\label{tab-obs_results}
\end{center}
\end{table}

\subsection{Subtracting the Lockman Window Spectrum}
\label{sec-lockman}
To measure accurately faint H$\alpha$ emission from high velocity
clouds, it would be a great benefit to locate a direction in the sky
where the Galactic H$\alpha$ emission was zero at all velocities.
Using this as the ``off'' direction would allow one to subtract the
sky spectrum and correct for other systematic uncertainties without
introducing the uncertainty caused by an unknown quantity of Galactic
emission from the ``off'' direction.  The Lockman Window (LW)
direction ($l$ = 150.48$^{\rm o}$, $b$ = 52.96$^{\rm o}$) appears to be
a good approximation to this ideal.  This direction has a famously low
H~I column density (N$_{H I}$ = 4.4 $\pm$ 0.5 $\times$ 10$^{19}$
cm$^{-2}$; Jahoda, Lockman, \& McCammon 1990) and
as a result is often used by X-ray astronomers and others interested
in looking at sources unobscured by the affects of neutral hydrogen
and the associated dust.  An important question is whether the Lockman
Window is similarly devoid of emission from ionized hydrogen.  Because
of baseline uncertainties in the WHAM spectra, we cannot claim that
there is zero H$\alpha$ emission from the Lockman Window direction;
however, it appears to be fainter in H$\alpha$ than any other
direction we have observed.  There is no evidence for H$\alpha$
emission from the Lockman Window except for a possible very weak
($\leq$ 0.02 R) high velocity component at v$_{lsr}$ $\simeq$ -130 km
s$^{-1}$.  We will eventually be able either to detect H$\alpha$ or
put very low upper limits on its intensity by using the shifts in
velocity caused by the earth's orbital motion to distinguish between
any actual emission and baseline irregularities.  The use of the
Lockman Window as an ``off'' direction therefore provides the best
indication of the amount of H$\alpha$ from both the high and the
intermediate velocity gas from the other directions, without the
biases caused by emission from an ``off'' direction containing
emission at lower (intermediate) velocities.  This lower velocity
emission in other ``off'' directions produces distortions in the
baseline of the resulting ``on -- off'' spectrum on the red side of
the HVC emission component (e.g., Fig~\ref{fig-hvc_c}).

\begin{figure}
  \centerline{\hbox{
      \epsfysize=2.7in \epsfbox{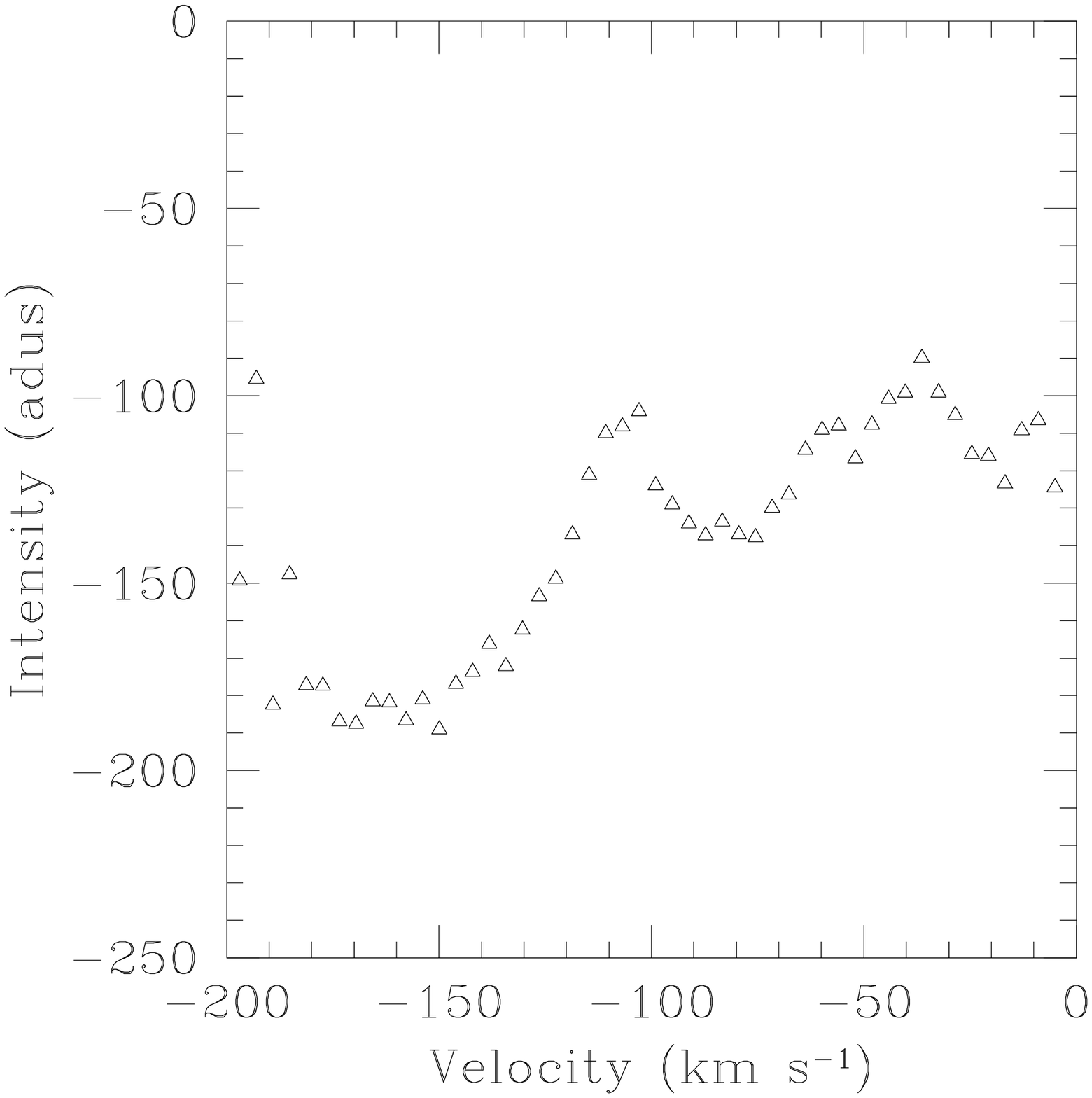}
      \epsfysize=2.7in \epsfbox{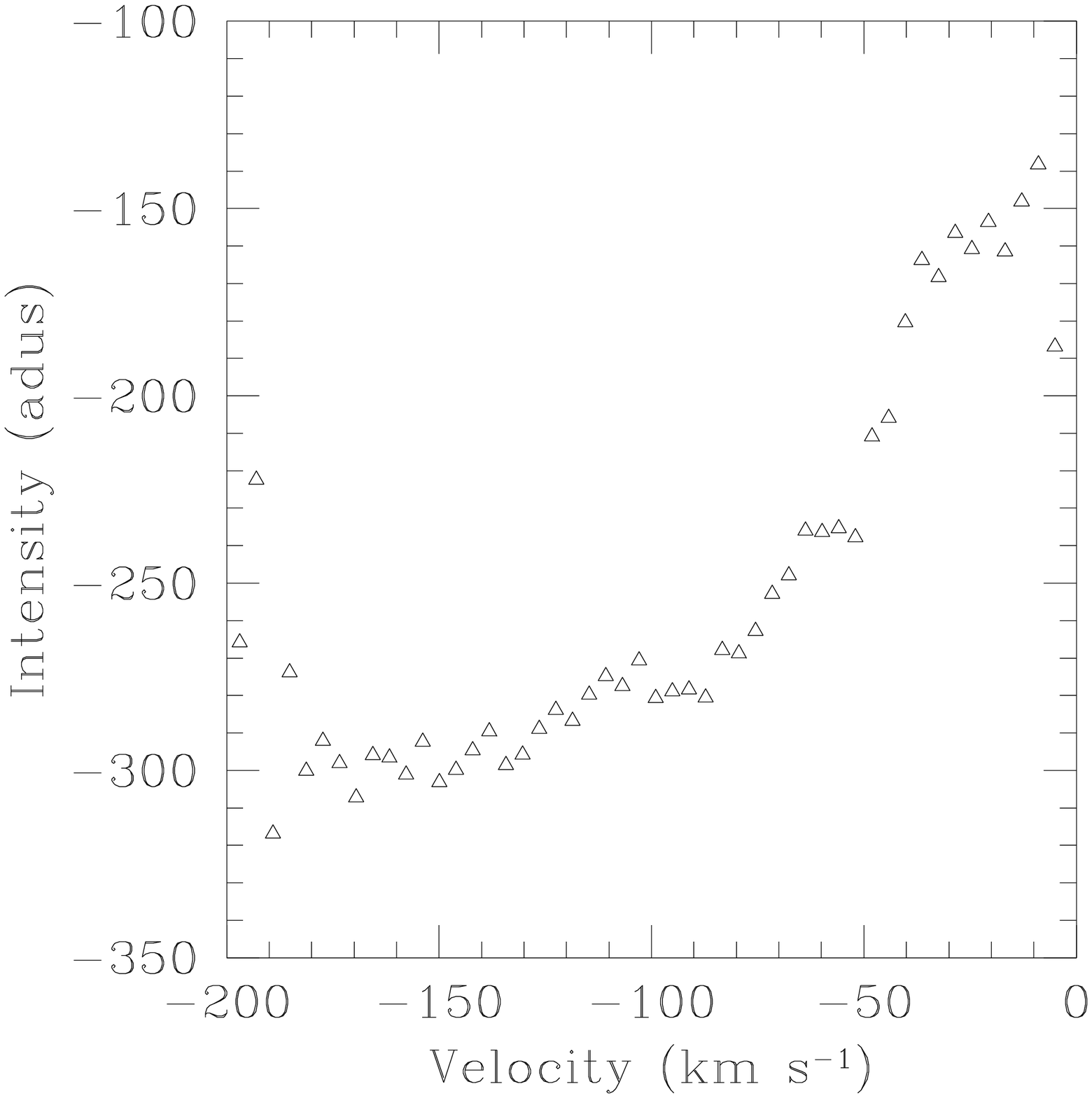}}
    }
  \caption{
    H$\alpha$ from the M I cloud: 1a - LW (left) and 
    4a - LW (right).  Both spectra are normalized to 900s exposure.}
\label{lockman}
\end{figure}

Figure~\ref{lockman} shows the 1a - LW difference spectrum.  In this
case, the emission component from the high velocity gas in the M~I
cloud is clearly revealed, just as it is revealed by the 1a -- 4a
difference spectrum in Figure~\ref{fig-hvcs}. However, in addition to
the HVC emission, H$\alpha$ is also seen at lower negative velocities,
identified with the intermediate velocity (-50 km~s$^{-1}$) H~I in
this direction.  There is perhaps also significant emission at even
lower velocities near the LSR, but this is less certain because
uncertainties associated with the subtraction of the geocoronal
H$\alpha$ line dominate at velocities more positive than -50 km
s$^{-1}$ (Tufte 1997).  Figure~\ref{lockman} also
shows the 4a - LW spectrum.  The intermediate velocity emission from
this direction is very similar to that seen in 1a, and this explains
why the baseline is so well behaved in the 1a--4a spectrum.
Differences in the amount of emission from intermediate velocity gas
between ``on'' and ``off'' HVC directions are likely responsible for
the baseline irregularities in Fig~\ref{fig-hvc_c}.

\section{The Relationship Between H$\alpha$ and 21-cm Emission}
\label{sec-disc}

\begin{figure}[tb]
  \centerline{\hbox{
      \epsfysize=2.7in \epsfbox{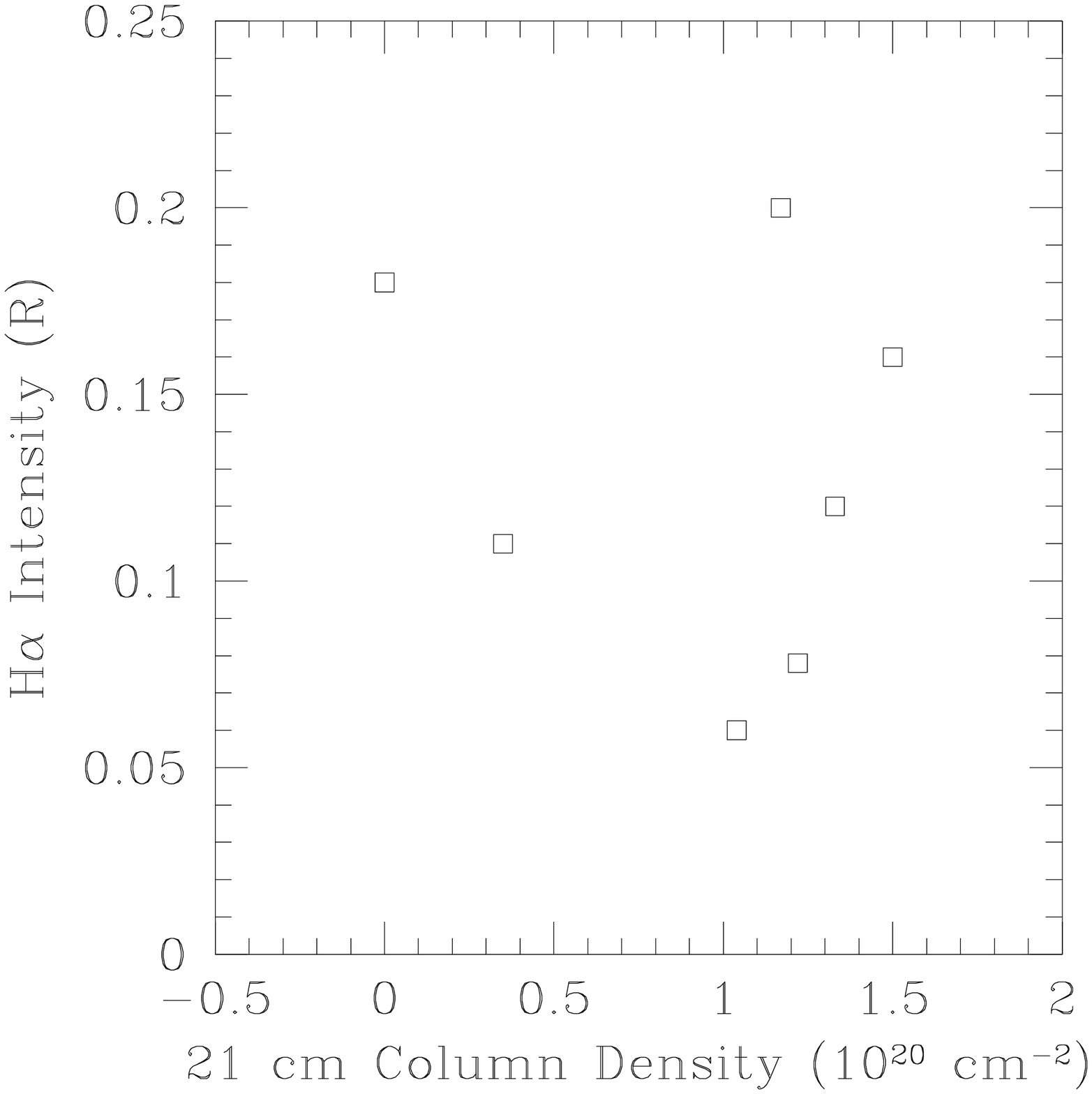}
      \epsfysize=2.7in \epsfbox{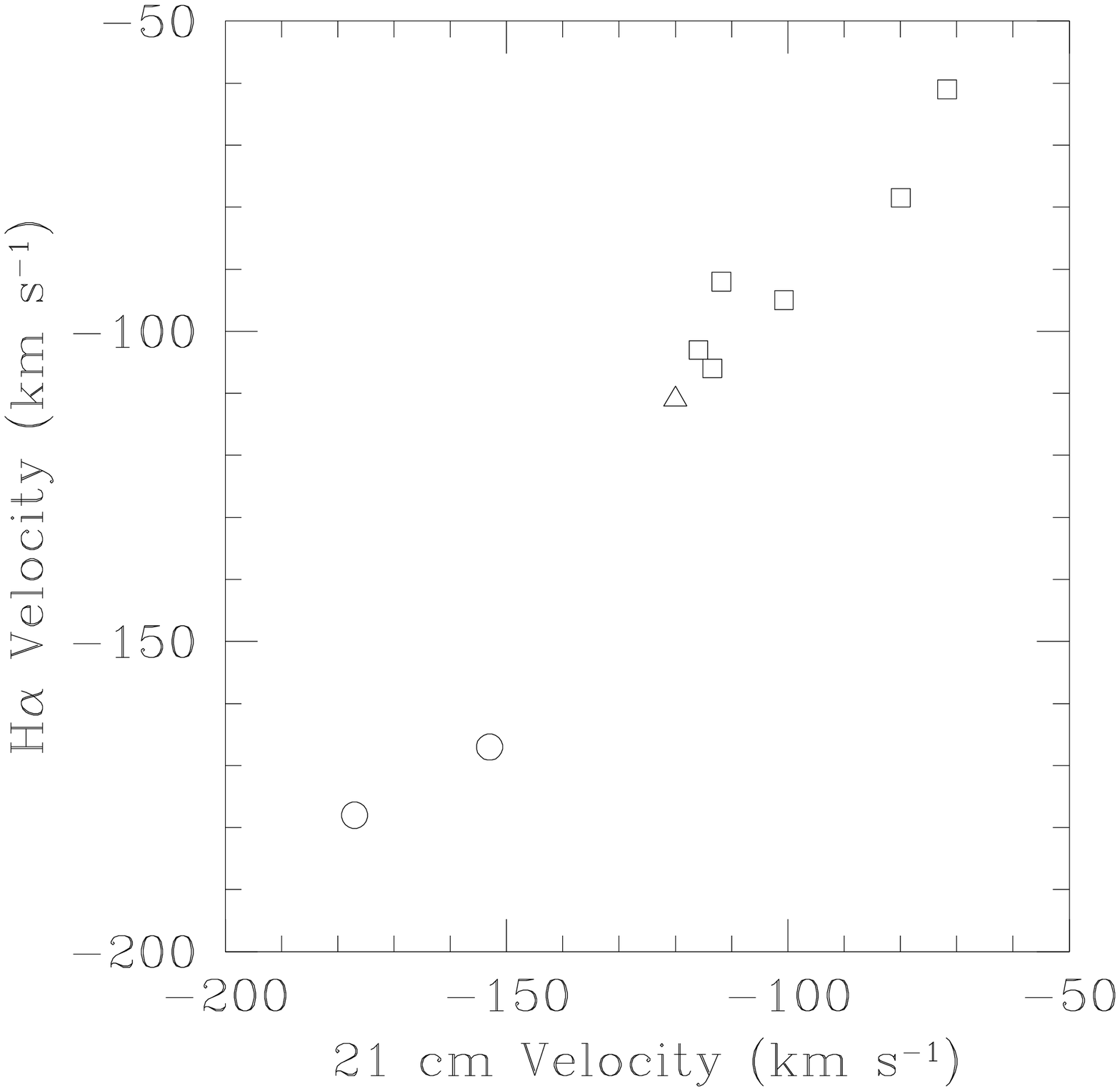}}
    }
  \caption{Comparison of H$\alpha$ and 21-cm emissions.  Left: 
     H$\alpha$ intensity versus 21-cm column density for the M Complex,
     Right: H$\alpha$ velocity versus 21-cm velocity.  Squares = M Complex,
     Circles = A Complex, Triangle = C Complex}
\label{hvc_corr}
\end{figure}

Since the H$\alpha$ intensity traces ionized hydrogen and the 21-cm
intensity traces neutral hydrogen, it is natural to wonder how these
two observables are related to each other.  For the WHAM observation
directions, we find that in every direction with significant high
velocity 21-cm emission (nine), an associated H$\alpha$ component was
detected.  Also, except for one observation direction (6a on
Fig.~\ref{fig4-map_m}), which is just off the edge of the M~I cloud,
there is no clear evidence of H$\alpha$ in any of the directions
without appreciable 21-cm emission.  This close correspondence,
however, does not imply that the intensities are well correlated, as
is demonstrated in the left panel of Figure~\ref{hvc_corr}, which
shows no clear correlation between the H$\alpha$ and 21-cm intensities
for the M complex.  The radial velocities, on the other hand, do show
a close correlation, as can be seen in the right panel of
Figure~\ref{hvc_corr} where the H$\alpha$ velocities are plotted
versus the 21-cm velocities for the M, A, and C complexes (each
complex is plotted with a different symbol).  This correlation is
strong evidence that the H$\alpha$ emission is associated with the
high velocity neutral hydrogen.

There is some indication that the ionized gas envelopes the neutral
gas.  Direction 6a (see Fig.~\ref{fig4-map_m}) at the edge of the M~I
cloud is brighter than direction 1a, which is on the M~I cloud where
the 21-cm is significantly brighter.  The faintness of the H$\alpha$
in direction 7a, however, argues that the H~II envelope, if present,
is not very thick.  Weiner \& Williams (1996) also
find an edge brightening effect in their Magellanic Stream data.  Of
their seven observation directions, the three that are brightest in
H$\alpha$ are on the edge of H~I clouds.  They further claim that the
association is with ``leading'' edges as defined by the inferred
motion direction of the gas; however, the evidence for this is not
strong as there is only one direction that could be called a
``trailing'' edge, and this direction has an intermediate H$\alpha$
brightness.  It will be important in future observations to
distinguish between edge brightening and leading edge brightening,
since edge brightening is naturally explained in a photoionization
scenario, whereas leading edge brightening is more naturally explained
by a collisional process.

The HVCs presented here are at high Galactic latitude, and therefore
interstellar extinction can probably be ignored.  In this case, the
H$\alpha$ intensity is directly related to the emission measure (EM
$\equiv \int n_{\rm e}^{2} dl$) through the cloud and gives
information on the column density and electron density of the ionized
hydrogen.  Consider the 21-cm enhancement in the M~I cloud centered
near direction 1a in Figure~\ref{fig4-map_m}, as an example.  It has
an H~I column density N$_{HI}$ $\simeq$ 1.2 $\times$ 10$^{20}$
cm$^{-2}$ and a diameter of about 1.5$^{\rm o}$, corresponding to 50
pc at a distance of 2~kpc and indicating a density of neutral hydrogen
n$_{\rm{H~I}}$ $\simeq$ 1.0 cm$^{-3}$ and a mass of about 1600
M$_{\odot}$ for the cloud.  The measured H$\alpha$ intensity
I$_{\alpha}$ = 0.08 R implies an emission measure EM = 0.18 cm$^{-6}$
pc, assuming a temperature for the H~II of 8000 K.  This leads to an
electron density n$_{\rm e} \simeq$ 0.06 f$^{-1/2}$ cm$^{-3}$ and a
column density of ionized hydrogen N$_{\rm{H~II}} \simeq 1 \times
10^{19}$ f$^{1/2}$ cm$^{-2}$, where f is the filling fraction of the
H~II along the line-of-sight through the cloud.  Since f $\leq$ 1 and
there is little evidence that the H~II associated with the cloud
extends significantly beyond the H~I portion of the cloud, it appears
that toward direction 1a the HVC is primarily neutral with N$_{H II}$
/ N$_{H I}$ $\leq$ 0.08.  On the other hand, if we consider direction
6a, which intersects the edge of the cloud (see
Fig.~\ref{fig4-map_m}), the H~I column density is less than 2 $\times$
10$^{19}$ cm$^{-2}$, whereas the H$_{\alpha}$ is twice the intensity
of 1a.  This indicates that there may be as much or more H~II as H~I
along the line of sight at the edge of the H~I enhancement.  Both of
these results appear to be consistent with most of the H~II being
located near the outer surface of the cloud rather than mixed with the
H~I.

\section{Implications of the H$\alpha$ Emission}
An important question is the source of ionization for the HVCs.  If
the H$\alpha$ emission arises from photoionization, then the intensity
is directly related to the incident Lyman continuum flux F$_{LC}$ in
the Galactic halo.  Since the H~I cloud is optically thick in the
Lyman continuum and optically thin to H$\alpha$ photons, each Lyman
continuum photon incident on the cloud will ionize a hydrogen atom and
each hydrogen recombination will produce on the average 0.46 H$\alpha$
photons (Martin 1988; Pengally 1964; case B, T $\sim$ 10$^{4}$).  This
leads to the relation
\[ F_{LC}= 2.1 \times 10^{5} \left(\frac{I_{\alpha}}{0.1 R}\right) 
 {\rm \:\:photons\:cm}^{-2} \rm{s}^{-1}. \] If, on the other hand, the
ionization is due to a shock arising from the interaction of the HVC
with ambient gas in the Galactic halo, the H$\alpha$ intensity
I$_{\alpha}$ provides information about the density and temperature of
the halo gas.  If the ambient gas is sufficiently cool (T $\leq$
10$^{6}$) for strong shocks to occur, then shock models by Raymond
(1979) relate the face-on H$\alpha$ surface brightness
(I$_\alpha$)$_\perp$ to the halo gas density. Namely,
\[ (I_{\alpha})_\perp \simeq 6.5 n_{\rm o} 
\left(\frac{V_{\rm s}}{100}\right)^{1.7} R, \] where n$_{\rm o}$ is
the density of the preshocked gas.  This power law is a fit to the
predicted H$\alpha$ intensities for shock velocities between 50
km~s$^{-1}$ and 140 km s$^{-1}$ presented in Raymond (1979).  For the
A~IV cloud, for example, where V$_{s}$ $\simeq$ 178 km~s$^{-1}$, and
(I$_{\alpha}$)$_\perp$ $\simeq$ 0.1 R, this would imply n$_{\rm o}$
$\leq$ 6 $\times$ 10$^{-3}$ cm$^{-3}$, which is an upper limit on
n$_{\rm o}$ because a nonperpendicular sightline will increase the
observed I$_\alpha$ for a given n$_{\rm o}$ and V$_{s}$.

The fact that the observed intensities are all around 0.1 R
(Table~\ref{tab-obs_results}) may be more easily explained in a
photoionization scenario because the various clouds could all be
bathed in the same ionizing flux (see Bland-Hawthorn \& Maloney 1998)
independent of their velocity.  It is more difficult to explain the
uniformity of the H$\alpha$ intensities in a shock excitation
scenario, since in that case the expected H$\alpha$ intensity varies
strongly with both the shock velocity and the ambient density.
Another weak piece of evidence for photoionization comes from the
fairly narrow, albeit poorly measured, H$\alpha$ line widths, which
are similar to the widths observed in photoionized H~II regions
surrounding O stars (Reynolds 1988).  A more careful map of
I$_{\alpha}$ around the edge of the clouds could perhaps distinguish
between the two ionization mechanisms.  The source of ionization could
also be explored through measurements of various emission line ratios.
With an ambient density n$_{\rm o}$ = 1 cm$^{-3}$ and a shock velocity
V$_{s}$ = 100 km s$^{-1}$, some of the line ratios predicted by
Raymond (1979) are [S~II] $\lambda$6716 / H$\alpha$ = 0.18, [N~II]
$\lambda$6584+$\lambda6548$ / H$\alpha$ = 0.51, and [O~III]
$\lambda$5007+$\lambda$4959 / H$\alpha$ = 1.9.  If this model, when
extended to lower densities, retains the prediction of very bright
[O~III] emission, then the $\lambda$5007 line could be an important
discriminator between shock ionization and photoionization.  Such
emission lines could also give further information on the metal
abundances in the HVCs, which have been previously investigated by Lu
et al. 1994 and Lu et al. 1998, for example, through absorption line
techniques (also see Wakker et al., this proceedings).

If indeed the detected H$\alpha$ emission results from photoionized
gas, then such measurements can shed light on more general questions
concerning the gaseous Galactic halo and its connection to processes
occurring in the disk.  It has frequently been proposed, for example,
that the source of the ionization in the extended warm ionized
component of the interstellar medium (the WIM) is O stars in the plane
of the Galaxy (e.g., Domg\"{o}rgen \& Mathis 1994), although other,
more exotic sources have also been proposed (e.g., Melott et al. 1988;
Sciama 1990).  However, for the O star idea to work, very special
arrangements of the H~I gas must exist to allow photons to travel from
the Galactic plane, where the vast majority of O stars are located, to
gas high above the plane in the thick WIM layer (for specific models
that have been constructed to explore this scenario in detail, see
Miller \& Cox 1993 and Dove \& Shull 1994).  If the H~I gas is
arranged to be porous to Lyman Continuum photons, then a significant
fraction of the photons should leak out of the disk and into the halo.
The H$\alpha$ intensity towards H~I clouds in the Galactic halo
constrains the degree to which this is occurring, and in particular,
the H$\alpha$ intensity towards H~I clouds at various distances from
the Galactic plane can reveal the Lyman continuum flux as a function
of height above the plane.  This idea and its implications have been
explored from a theoretical stand point by Bland-Hawthorn \& Maloney
(1998).  The H$\alpha$ intensities measured for the HVCs are about a
factor of 10 lower than typical H$\alpha$ intensities from WIM gas at
high Galactic latitudes (e.g., Reynolds et al. 1995; Haffner et
al. 1998), implying that if the ionizing photons originate near the
Galactic midplane, 90\% of the photons reaching the WIM are absorbed
before reaching the even greater heights of the HVCs in the Galactic
halo.  Note that the measured H$\alpha$ intensities for the HVCs are
an order of magnitude (or more) above that expected from the
metagalactic ionizing flux (Ferrera \& Field 1994).

In the near future, there should be much progress in this emerging
area of study.  Entire HVC complexes will be mapped in H$\alpha$,
other diagnostic lines will be measured revealing information about
the source of ionization, and comparative studies between H$\alpha$,
21-cm, X-ray, and absorption line measurements will provide clues
about the relationship between the various gas phases in the Galactic
halo.

We thank N. R. Hausen for her valuable contributions.  Observations
with the WHAM facility have been supported by the National Science
Foundation through grants AST 9619424 and AST 9122701.

\end{document}